# Magnetic phase diagram and first principles study of $Pb_3TeCo_3V_2O_{14}$

M.M. Markina,[1] B.V. Mill,[1] E.A. Zvereva,[1] A.V. Ushakov,[2]
S.V. Streltsov,[2,3] and A.N. Vasiliev[1,3]

[1]Physics Faculty, M.V. Lomonosov Moscow State University, Moscow 119991, Russia
[2]Institute of Metal Physics, S. Kovalevskoy Street 18, Ekaterinburg 620990, Russia
[3]Theoretical Physics and Applied Mathematics Department, Ural Federal University, Mira Street 19, Ekaterinburg 620002, Russia

An antiferromagnetic ordering in $Pb_3TeCo_3V_2O_{14}$ takes place through formation of short range correlation regime with $T^* \sim 10.5$ K and succession of second order phase transition at $T_{N1} = 8.9$ K and first order phase transition at $T_{N2} = 6.3$ K. An external magnetic field rapidly destroys magnetic structure at $T < T_{N2}$ and influences the magnetic order at $T_{N2} < T < T_{N1}$ resulting in complex magnetic phase diagram of $Pb_3TeCo_3V_2O_{14}$ as derived from magnetization and specific heat measurements. The first principles calculations indicate that in variance with layered crystal structure the magnetic subsystem of $Pb_3TeCo_3V_2O_{14}$ is quasi-one-dimensional and highly unusual consisting of weakly coupled triangular tubes.

**Introduction**

$Pb_3TeCo_3V_2O_{14}$ is a representative of large group of compounds with a common formula $A^{2+}_3Te^{6+}M^{2+}_3X^{5+}_2O_{14}$ (A = Pb, Ba, Sr; M = Zn, Mg, Co, Mn, Cu, Cd; X = P, As, V) and $Pb_3WZn_3P_2O_{14}$ [1] with the $Ca_3Ga_2Ge_4O_{14}$ crystal structure (sp. gr. $P321$ [2]). It is a close analogue of dugganite mineral (ideal formula $Pb_3TeZn_3As_2O_{14}$) whose crystal structure [3] is the same as $Ca_3Ga_2Ge_4O_{14}$ and numerous compounds of so-called langasite $La_3Ga_5SiO_{14}$ family. Recent investigations of $Pb_3TeCo_3V_2O_{14}$ indicate that dugganite group compounds possess a complicated crystal structure based on $Ca_3Ga_2Ge_4O_{14}$ one with small monoclinic distortion (sp. gr. $P121$) and multiplicated cell parameters.[4-6]

The presence of magnetic $Co^{2+}$ and $Mn^{2+}$ ions in the structure add new, magnetic dimension to langasite's. Magnetic properties in these compounds are related to such substitutions since another transition metal in the structure, vanadium, is pentavalent and do not carry magnetic moment. It is also possible that the members of this family can exhibit the multiferroic properties. Depending on the mechanism of inversion symmetry breaking the multiferroics i.e. materials maintaining both electric and magnetic spontaneous polarizations are classified similar to proper and improper ferroelectrics.[7] In the proper multiferroics covalent bonding or polarization of the ions having non-magnetic lone pairs of s-electrons, e.g. Bi or Pb, results in ferroelectric distorsions.[8] The improper ferroelectricity can arise as a secondary effect driven by complex magnetic ordering.[9] Both these mechanisms could be relevant in langasites due to $Pb^{2+}$ single ion property and tricky magnetic response of $Pb_3TeCo_3V_2O_{14}$.

Recently, the magnetic and structural properties of $Pb_3TeCo_3V_2O_{14}$ were the subject of intense studies.[4-6] It was assumed initially, that this compound crystallizes in trigonal unit cell with noncentrosymmetric $P321$ space group, similar to langasite. At lowering temperature, $Pb_3TeCo_3V_2O_{14}$ orders magnetically in two steps at $T_{N1} \sim 9$ K and $T_{N2} \sim 6$ K. Both these transitions are reflected in dielectric permittivity signaling the coupling of magnetic and electric subsystems. It was claimed that the spins adopt incommensurate magnetic structure with propagation vector $\boldsymbol{q}_1 = (0.752, 0, 0.5)$ at $T_{N1}$ and reorder into a commensurate magnetic structure with propagation vector $\boldsymbol{q}_2 = (5/6, 5/6, 1/2)$ at $T_{N2}$.[4] Then, basing on synchrotron data a monoclinic superstructure with space group P121 was established yielding a very large supercell.[5] Finally, precise X-ray diffraction and neutron scattering measurements were carried out resulting in new magnetic structure solutions with propagation vectors $\boldsymbol{q}_1 = (½, 0, -½)$ at $T_{N2} < T < T_{N1}$ and $\boldsymbol{q}_2 = (½, ½, -½)$ at $T < T_{N2}$.[6] It was shown, that the low temperature magnetic structure of every Te-containing member of the langasite family, i.e. $Pb_3TeCo_3As_2O_{14}$, $Pb_3TeCo_3P_2O_{14}$, and $Pb_3TeCo_3V_2O_{14}$, is quite sensitive to magnetic field.[5]

The aim of the present work is to establish magnetic phase diagram, treat rather weak magnetoelectric effect and estimate intra- and interlayer magnetic exchange interaction parameters in $Pb_3TeCo_3V_2O_{14}$ by the first principles calculations.

**Experimental**

*Sample preparation and crystal structure.* Polycrystalline samples of $Pb_3TeCo_3V_2O_{14}$ were obtained through standard solid-state reaction method with starting chemicals PbO, $TeO_2$, $Co_3O_4$, and $V_2O_5$ of high purity. Stoichiometric by metal composition mixture was ground in the agate vibromill Pulverisette-0 (Fritsch), pressed into bars, and sintered in air at 850°C. Similar procedure was employed for synthesis at 950°C of the reference for specific heat measurements $Pb_3TeZn_3As_2O_{14}$. The phase purity of the samples was controlled by the X-ray diffraction at room temperature. The impurities content was found to be about 1 % by mass. The crystal structure of $Pb_3TeCo_3V_2O_{14}$, as defined in Ref. 6 is shown in Fig. 1. Both large $CoO_4$ and small $VO_4$ tetrahedra constitute slightly corrugated layers with triangular motif in the arrangement of magnetically active $Co^{2+}$ ions. Another layer is organized by $PbO_{10}$ decahedra and $TeO_6$ octahedra. The $Pb^{2+}$ ions are arranged in zig-zag manner within channels of magnetically passive layers. The monoclinic $P121$ cell alters the equilateral trimer of the $P321$ subcell into six isosceles trimers per unit cell. Within the magnetically active layers the cobalt ions are connected through either two oxygens belonging to $VO_4$ unit (large cobalt triangle) or two oxygens belonging to $TeO_6$ unit (small cobalt triangle). The interlayer magnetic interaction involves $TeO_6$ octahedra, as discussed below.

*Electron spin resonance.* Low-frequency electron spin resonance (ESR) studies were carried out using an X-band ESR spectrometer CMS 8400 (ADANI) ($f \approx 9.4$ GHz, $B \leq 0.7$ T) equipped with a low temperature mount, operating in the range $T = 5 - 300$ K. The effective g-factors of our samples have been calculated with respect to a BDPA (a,g-bisdiphenyline-b-phenylallyl) reference sample with $g$-factor 2.00359. The evolution of the X-band ESR spectra with temperature for a powder sample of $Pb_3TeCo_3V_2O_{14}$ is shown in Fig. 2 (lower panel). The weak anisotropic signal observed is characteristic of high spin $Co^{2+}$ in tetrahedral coordination. No hyperfine structure arising from an electron-nucleus interaction for the $^{59}Co$ isotope (100% abundance, $I = 7/2$) is observed in the spectrum. The ground state of $Co^{2+}$ in a tetrahedral crystal field is fourfold spin degenerate, as in the first order the $e_g$ subshell is completely filled and the $t_{2g}$ subshell is half-filled giving $S = 3/2$ with no orbital degeneracy. In second order, however, the on-site Coulomb and exchange interactions (multiplet effects) mix a certain amount of $e_g$ electrons into the $t_{2g}$ shell and restore the orbital momentum partially. This mixing depends on the relative size of the ligand field splitting and the spin-orbit coupling and does not split the ground state. But its energy is lowered and the total magnetic moment is enhanced by the orbital momentum. For $S = 3/2$ spin system (HS $Co^{2+}$, $d^7$), the zero field perturbation splits the energy levels into two doublets, $|\pm 1/2\rangle$ and $|\pm 3/2\rangle$. Thus, the nature of the spectrum could be assigned to the transitions between the ground state Kramers doublet with anisotropic $g$-factor, which is assumed to originate from the involvement of an excited doublet with an effective spin of 3/2.[10] The powder X-band ESR spectrum is very weak and overlapping in nature, thereby eliminating any possibility of precise determination of the individual transverse $g_\perp$ and longitudinal $g_\parallel$ components of the $g$-tensor of the powder sample. However, the estimated from Lorenzian fit $g$ values $g_\perp \approx 2.13$ and $g_\parallel \approx 3.10$ are in reasonable agreement with the experimental values observed for other compounds with $Co^{2+}$ ions in tetrahedral coordination.[11-13] Evidently, the ESR line is characterized by practically temperature-independent effective $g$-factors with an average value $g = (2g_\perp + g_\parallel)/3 = 2.45 \pm 0.05$. A rough estimation for an exchange anisotropy according to expression $J_\perp/J_\parallel = g_\perp/g_\parallel$ [14] yields a value $J_\perp/J_\parallel \sim 0.7$ indicating moderate anisotropy in $Pb_3TeCo_3V_2O_{14}$. With decreasing temperature the line narrows slightly probably owing to decreasing role of the spin-lattice relaxation mechanism. Typical for antiferromagnets, the so-called critical broadening due to slowing down of the spin-spin correlations upon approaching the Neel temperature from above was not observed.

*Magnetometry.* Magnetic measurements were conducted in a "Quantum Design" Physical Property Measurements System PPMS-9T. The temperature dependences of the magnetization were measured under various magnetic fields up to $B = 9$ T in the temperature range 2 – 300 K. In addition, the isothermal magnetization curves were obtained in external fields up to 9 T at various constant temperatures in the range 2 – 10 K after cooling the sample in zero magnetic field. The temperature dependences of the reduced magnetization $M/B$ taken at heating in various magnetic fields in the range 0.1 – 9 T are shown in Fig. 3. When measured at lowest magnetic field $B = 0.1$ T the magnetization passes through broad maximum at $T^* \sim 10.5$ K and drops down essentially twice at low temperatures. The latter fact itself suggests that the magnetic anisotropy in $Pb_3TeCo_3V_2O_{14}$ is of the easy-plane type. The broad maximum is not a signature of the phase transition but relates to formation of the short-range correlation regime. It is seen routinely in low-dimensional and/or frustrated magnetic systems. Despite seemingly two-dimensional and frustrated geometry of the magnetic subsystem neither of these factors prevent formation of the long-range order in $Pb_3TeCo_3V_2O_{14}$. As shown in the Inset to Fig. 3, the magnetic ordering takes place in two steps marked by singularities in d($M/B$)/d$T$ vs. $T$ curves. The λ-type peak marks the phase transition at $T_{N1} = 8.9$ K, while the phase transition at $T_{N2} = 6.3$ K is marked by a sharp spike. Just the shapes of these singularities indicate different nature of the transitions, i.e. second order transition at $T_{N1}$ and first order transition at $T_{N2}$. Therefore, merely the closeness of $T^*$ and $T_{N1}$ indicates that in variance with crystal structure the magnetic subsystem in $Pb_3TeCo_3V_2O_{14}$ is neither quite low-dimensional nor heavily frustrated. This conclusion is supported by the measurements in wide temperature range (not shown) and results of *ab initio* calculation presented below. At high temperatures, the magnetic susceptibility of the title compound follows the Curie-Weiss law with addition of the temperature independent term:

$$\chi = \frac{M}{B} = \chi_0 + \frac{C}{T-\Theta} = \chi_0 + \frac{g^2\mu_B^2 S(S+1)}{3k_B(T-\Theta)} \tag{1}$$

The Curie constant $C = 8$ emu/mol K, as defined from 200 – 300 K range, corresponds to magnetic moments of $Co^{2+}$ ions ($n = 3$ – number in the formula unit) in the high-spin state $S = 3/2$ with addition of some unquenched orbital moment. Using averaged $g$ – factor $g = 2.45$ gives estimation of the Curie constant $C_{est} = 8.2$ emu/mol K in fair agreement with experimentally found value. The temperature independent term is positive $\chi_0 = 1.67 \times 10^{-3}$ being a summation of the paramagnetic van Vleck contribution of the cobalt ions[12] and individual ions Pascal's diamagnetic constants.[15] Since the Weiss temperature $\Theta = -25$ K one may expect modest frustration ratio of $|\Theta|/T_{N1} \sim 2.8$.

When measured at higher magnetic fields the signatures of the phase transitions shift to lower temperatures and broaden, especially that one at $T_{N2}$. The change in the shape of $T_{N2}$ singularity signifies that the order of this phase transition changes gradually under magnetic field from the first order to the second order.

The field dependences of magnetization in $Pb_3TeCo_3V_2O_{14}$ are shown in Fig. 4. At highest available magnetic field $B = 9$ T and lowest temperature $T = 2$ K, the magnetic moment $M \sim 4$ $\mu_B$/f.u. is still far from the estimated saturation value $M_{sat} = ng\mu_B S \sim 11$ $\mu_B$. The derivative d$M$/d$B$ vs. $B$ shown in the Inset to Fig. 4 indicates that there exist some singularities in the curve, i.e. jump at $B_1(T)$ and broad maximum at $B_2(T)$. These singularities shifting to lower magnetic fields and smearing with increasing temperature mark the field induced phase transitions. All differential curves from Figs. 3 and 4 are given in supporting information SI1, SI2.

*Dielectric permittivity.* Dielectric constant measurements in the range 4 – 20 K were made using an automated capacitance bridge Andeen-Hagerling AH-2700A operating in a frequency range $10^3$ - $10^5$ Hz. The Inset to Fig. 5 represents temperature dependence of the dielectric constant $\varepsilon/\varepsilon_0$ in $Pb_3TeCo_3V_2O_{14}$ taken at the frequency $10^4$ Hz. While the phase transition at $T_{N1}$ is hardly seen in this physical property, the phase transition at $T_{N2}$ is marked by a sharp drop $\Delta\varepsilon/\varepsilon_0 \sim 2\times 10^{-4}$. No strong frequency dependence was found at the phase transition singularities in the temperature range measured.

*Calorimetry.* The specific heat measurements were carried out by a relaxation method using a Quantum Design PPMS system. The plate shaped sample of 0.2 mm in thickness and ~ 6 mg in mass was obtained by compressing the polycrystalline powder. Data were collected at zero magnetic field in the temperature range 2 – 120 K and under several applied fields up to 9 T in the range 2 – 16 K. The temperature dependences of specific heat $C(T)$ in $Pb_3TeCo_3V_2O_{14}$ and $Pb_3TeZn_3As_2O_{14}$ are shown in Fig. 5. The phase transitions at $T_{N1}$ and $T_{N2}$ are clearly seen as λ-type anomaly and sharp spike, respectively. A special scaling procedure based on comparison with the reference compound as described in Ref. 16 allows the magnetic specific heat $C_{magn}$ and the magnetic entropy $S_{magn}$. The temperature dependence of the latter entity is shown in the lower Inset to Fig. 5. Evidently, significant part of the magnetic entropy is released well above the magnetic ordering temperatures, but the overall release is in good correspondence with the standard estimation $S_{magn} = nR\ln(2S+1) = 34.6$ J/mol K.

The temperature dependences of specific heat $C(T)$ taken at various magnetic fields in $Pb_3TeCo_3V_2O_{14}$ are shown in Fig. 6. As is the case of $dM/dB$ vs. T dependences, the application of magnetic field rapidly suppresses sharp singularity at $T_{N2}$ shifting both anomalies to low temperatures. When dealing with specific heat measurements done at various magnetic fields we defined the transition temperature through positions of peaks at $C$ vs. $T$ curves. It is shown also in supporting information SI3. The correspondence of the singularities in Fisher's specific heat $dM/dT$ vs. $T$ with those in heat capacity $C(T)$ is apparent.

*Phase diagram.* The data collected in $M(T)$, $M(B)$ and $C(T)$ measurements are used to establish the magnetic phase diagram of $Pb_3TeCo_3V_2O_{14}$, shown in Fig. 7. Since the results presented here were obtained on a polycrystalline sample, this phase diagram is equivalent to the average of the phase diagrams of a single crystal over all possible field directions. In case of pronounced magnetocrystalline anisotropy such a phase diagram corresponds to magnetic phase diagram in a magnetic field applied perpendicular to the hard axis of anisotropy, possibly with a slight droadening of the transitions. Therefore, the magnetic phase diagram obtained on a polycrystalline sample is that expected for a single crystal of $Pb_3TeCo_3V_2O_{14}$ in a magnetic field parallel to the easy axis or easy plane (depending on the nature of the anisotropy). [17] Note, in case of low-dimensional antiferromagnets the phase diagrams obtained on a polycrystalline samples are in good agreement with those obtained on a single crystals. [18-20]

Following Ref. 15, the phase I denote antiferromagnetically ordered state with the wavevector $q_2$ = (½,½, -½). The phase II corresponds to antiferromagnetically ordered state with the wavevector $q_1$ = (½, 0, -½). This phase transforms to magnetically induced phase III with finite magnetization, i.e. canted antiferromagnetic (ferrimagnetic) state, separated by a phase boundary from paramagnetic state denoted as the phase IV. The determination of this phase boundary through position of a broad maximum in $C$ vs. $T$ dependence is somewhat more reliable than from the $M(T)$ or $M(B)$ dependences since external magnetic field efficiently smears the anomaly in magnetization at transition between ferromagnetic and paramagnetic states. It is possible also that further investigations will lead to even more elaborated magnetic phase diagram in $Pb_3TeCo_3V_2O_{14}$ since the spin-flop-like transitions could be expected for both I and II phases. Since the measurements were conducted on a polycrystalline samples and the directions of hard and easy axes of anisotropy are not defined yet the arrows in Fig. 7 are just indicative.

**First principles calculations**

The calculation of the electronic and magnetic properties of $Pb_3TeCo_3V_2O_{14}$ were performed within the GGA+U approximation[17] in the framework of the pseudopotential method using Quantum Espresso code.[18] The exchange-correlation potential was taken in the form proposed by Perdew et. al.[19] The on-site Coulomb repulsion parameter for Co was chosen to be $U = 7$ eV.[20] Hund's rule coupling parameter ($J_H$) was taken to be $J_H = 0.9$ eV, so that $U_{eff} = U - J_H = 6.1$ eV. We also performed additional calculations for $U_{eff} = 5.6$ eV to show that the results obtained are stable with respect to small changes of *U*. The charge density cutoff equals 40 Ry. Integration in the course of self-consistency was performed over a mesh of 48 k-points in a full part of the Brillouin-zone.

The in-plane and out-of-plane exchange integrals were calculated fitting the total energy difference of several magnetic configurations (shown in Fig. 9) to the Heisenberg model written as

$$H = \sum_{ij} J_{ij} \vec{S}_i \vec{S}_j, \qquad (2)$$

where the summation runs once over every pair of *i,j*. The supercell consisting of 12 Co ions was used for that. The atomic positions and lattice constants were taken from Ref. 4. The crystal structure presented in Ref. 5 and consisting of 138 atoms in the unit cell (which has to be at least doubled to calculate J is the c direction) is too large to be used for the exchange constants calculation.

The total and partial densities of states (DOS) for $Pb_3TeCo_3V_2O_{14}$ are presented in Fig. 8. In the GGA+U approximation $Pb_3TeCo_3V_2O_{14}$ is an insulator with the band gap ~ 2 eV. The valence band is formed mostly by the O 2p and Co 3d states. There are also V 3d states below the Fermi level. This is due to a strong hybridization between V 3d and O 2p states, which is especially large in the case of the high oxidation state like for the $V^{5+}$ ions. The magnetic moment on V ions is zero, while the spin moment on Co equals 2.6 $\mu_B$. This corresponds to 2+ oxidation state of Co with $3d^7$ electronic configuration. The deviation from ionic spin-only value of $3\mu_B$ expected for $Co^{2+}$ is related with the hybridization effects. This is typical for many transition metal oxides.[21,22]

The direct calculation of the exchange integrals in the ab plane shows that the magnetic coupling inside "small" Co triangles is ferromagnetic, $J_2$ = - 1.4 K. It is much larger than exchange interaction in "big" triangles, $J_1$ = - 0.2 K. However this system should not be considered as a net of weakly coupled triangles, since there is a strong magnetic coupling in the c direction, between triangles lying in neighboring Co-V plane. Moreover, this coupling is highly untrivial.

The exchange integrals along the c axis by itself is no that large, $J_c$ = 2.7 K, but there is a substantial coupling along diagonals as shown in Fig. 9, described by parameters $J_{d1}$ and $J_{d2}$. Unfortunately, the present choice of the supercell does not allow to evaluate $J_{d1}$ and $J_{d2}$ separately, while increased in the c direction supercell with 24 Co ions (i.e. 184 atoms of different types) is too large to be calculated. However, we found that $J_{d1} + J_{d2}$ = 4.3 K (antiferromagnetic). So that, these are the largest exchange couplings in the system. Together with ferromagnetic $J_c$ and $J_2$ they relieve frustration and make Co ions neighboring in the c direction to be antiferromagnetically ordered, which agrees with experimental findings.[5,6]

We repeated calculation of the exchange constants with slightly smaller $U_{eff}$ and found that with decrease of $U_{eff}$ on 0.5 eV both $J_{d1} + J_{d2}$ and $J_2$ increase in absolute values as expected for superexchange ($J_{d1} + J_{d2}$ = 5.3 K, $J_2$ = - 2.0 K), while $J_1$ and $J_c$ practically do not change ($J_1$ = 0.1 K, $J_c$ = 2.6 K).

Relatively strong exchange coupling inside small triangles $J_2$ one can explain by small Co–Co distance (3.52 Å). This is rather surprising that the interaction along triangular diagonals, $J_{d1}$ and $J_{d2}$ are comparable with $J_2$, since corresponding Co-Co bond length are almost twice as large (~ 6.29 Å).

This is obvious, that a strong exchange coupling on such distances in insulator may be obtained only via superexchange mechanism. There are two types of the atoms sitting along the diagonals: Te and O, which are arranged in two different ways. Along one of the diagonals Te and two O ions form nearly 180º bond (Fig. 10a), while along another diagonal this angle is close to 90º (Fig. 10b). One may speculate about two possible superexchange paths along these two diagonals. We note that these exchange constants were obtained for the crystal structure presented in Ref. 13, an additional distortions proposed in Ref. 5 may modify these results.

In the case of the 180° bond the superexchange interaction may occur between two $t_{2g}$ orbitals via two O $2p$ and one Te $p$ orbitals as shown in Fig. 10a. In the case of the 90° bond a more elegant way can be proposed. Since all $t_{2g}$ orbitals of the tetrahedraly coordinated Co$^{2+}$ ion with spin majority are half-filled one may construct $a_{1g} = (d_{xy} + d_{yz} + d_{zx})/\sqrt{3}$ orbital, which has a strong overlap with oxygen $p_\sigma = (p_x + p_y + p_z)/\sqrt{3}$ orbital.[23] The $p_\sigma$ orbitals on two different oxygen sites, forming 90° O-Te-O bond are directed to each other, so that the final superexchange process will include two $pd$ (from Co $a_{1g}$ to O $p_\sigma$) and one $pp$ hoppings (between two $p_\sigma$ orbitals) as shown in Fig. 10b.

The presence of the strong diagonal interlayer exchange coupling can be very important for the ferroelectric properties. Similar diagonal exchange interaction between triangular layers was shown to induce spontaneous electric polarization in FeRbMo$_2$O$_8$.[24] It's also worthwhile mentioning that strong exchange coupling between small triangles in different Co-V planes leads to the conclusion that Pb$_3$TeCo$_3$V$_2$O$_{14}$ can be considered as a set of the weakly coupled 1D objects – triangular tubes.

**Conclusion**

It was found, that an antiferromagnetic ordering in Pb$_3$TeCo$_3$V$_2$O$_{14}$ takes place through formation of short range correlation regime with $T^* \sim 10.5$ K and succession of second order phase transition at $T_{N1} = 8.9$ K and first order phase transition at $T_{N2} = 6.3$ K. An external magnetic field rapidly destroys magnetic structure at $T < T_{N2}$ and influences magnetic structure at $T_{N2} < T < T_{N1}$ resulting in complex magnetic phase diagram of Pb$_3$TeCo$_3$V$_2$O$_{14}$ as derived from magnetization and specific heat measurements. The results of the first principles calculations show that the magnetic lattice of Pb$_3$TeCo$_3$V$_2$O$_{14}$ is highly unusual. It can be described as a net of weakly coupled 1D objects: triangular columns, or tubes consisting on $S = 3/2$. This makes Pb$_3$TeCo$_3$V$_2$O$_{14}$ important model system to study magnetic properties on such a lattice. In addition, this compound represents interesting example of the system with the long-range superexchange interaction.

**Acknowledgments**

This work is supported by the Russian Foundation for Basic Research through RFFI-13-02-00374, RFFI-13-02-00050 grants, the Ministry of education and science of Russia through grant MK-3443.2013.2 and by Samsung via GRO program. The part of the calculations were performed on the "Uran" cluster of the IMM UB RAS.

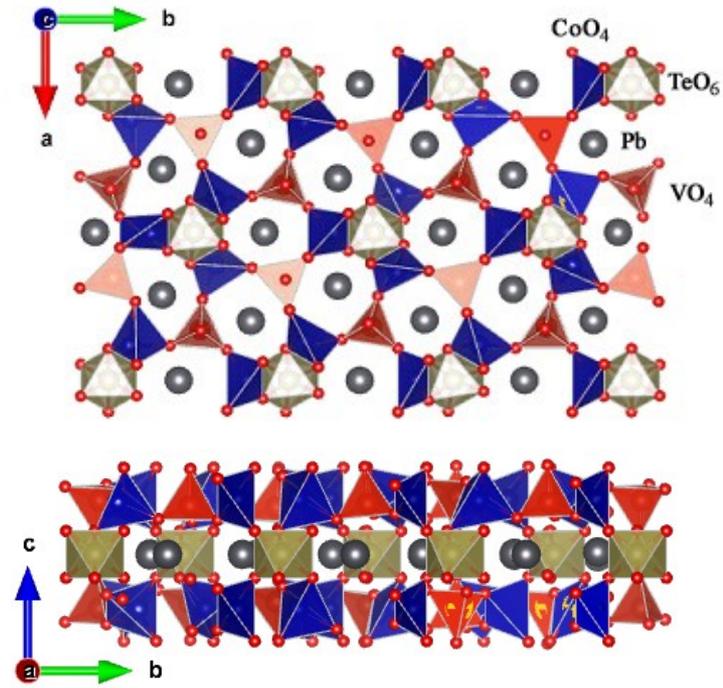

Fig. 1. The crystal structure of $Pb_3TeCo_3V_2O_{14}$.

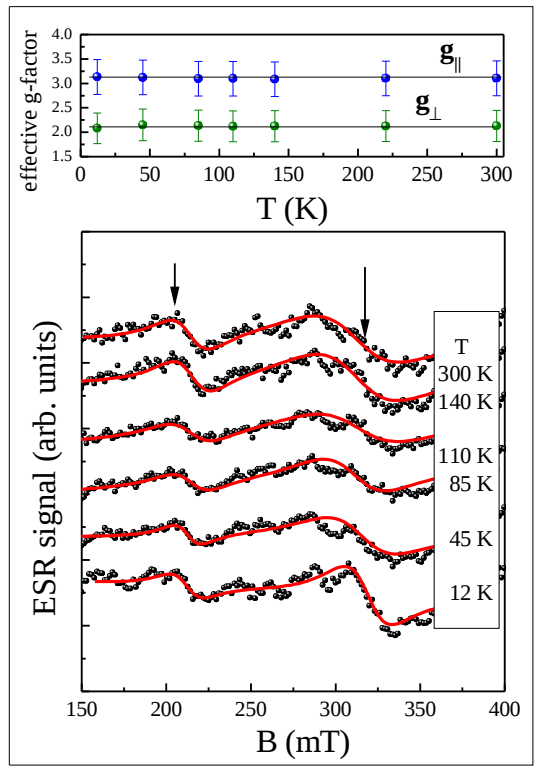

Fig. 2. Evolution of the first derivative of the ESR absorption line in $Pb_3TeCo_3V_2O_{14}$ with temperature: the black circles are experimental data, the red solid lines are approximations by sum of two Lorentzians (lower panel). The temperature dependence of the effective *g*-factors for two principal components of anisotropic ESR spectra (upper panel).

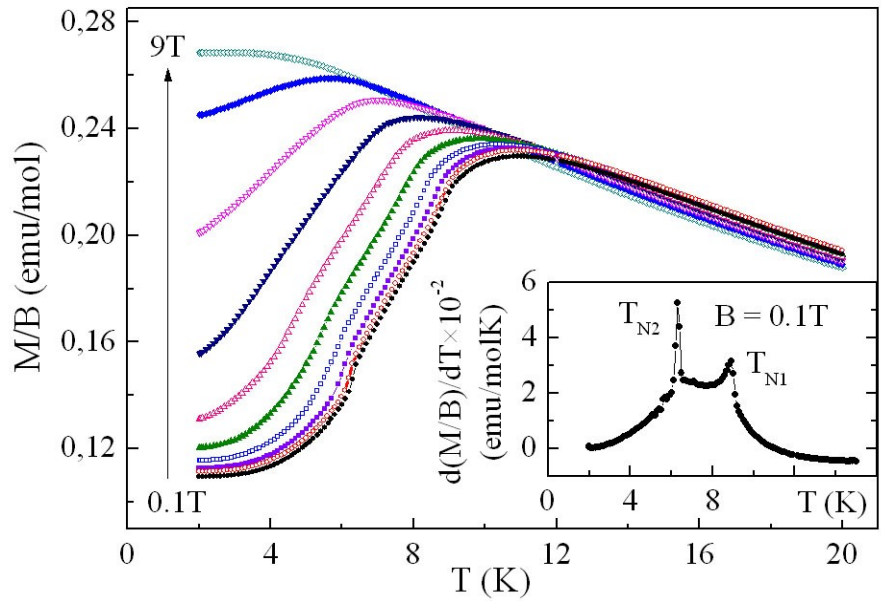

Fig. 3. The temperature dependences of reduced magnetization $M/B$ in $Pb_3TeCo_3V_2O_{14}$ taken at 0.1 K step in various magnetic fields in the range 0.1 ÷ 9 T. The Inset: the temperature derivative of the reduced magnetization taken at $B = 0.1$ T.

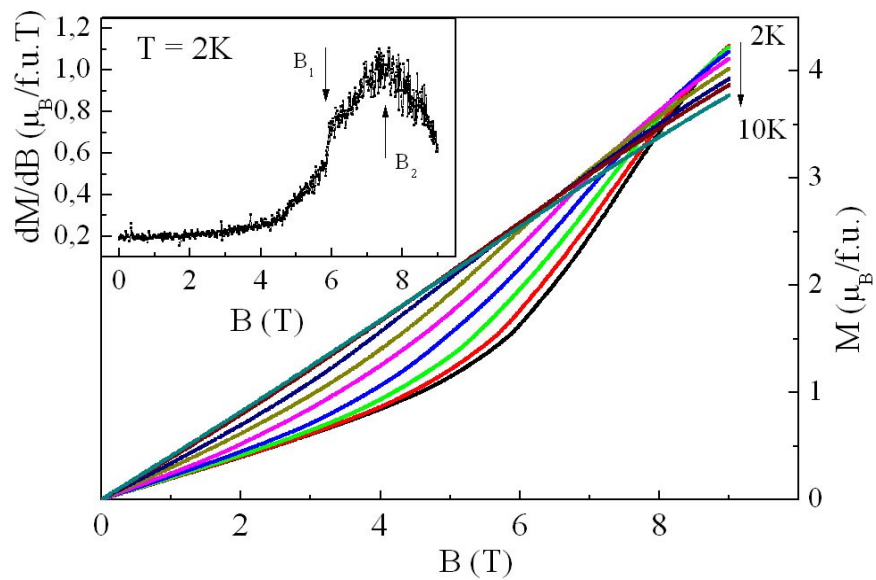

Fig. 4. The field dependences of magnetization in $Pb_3TeCo_3V_2O_{14}$ taken at various temperatures with 0.001 T step in the range 2 – 10 K. The Inset: the field derivative of the magnetization at $T = 2$ K.

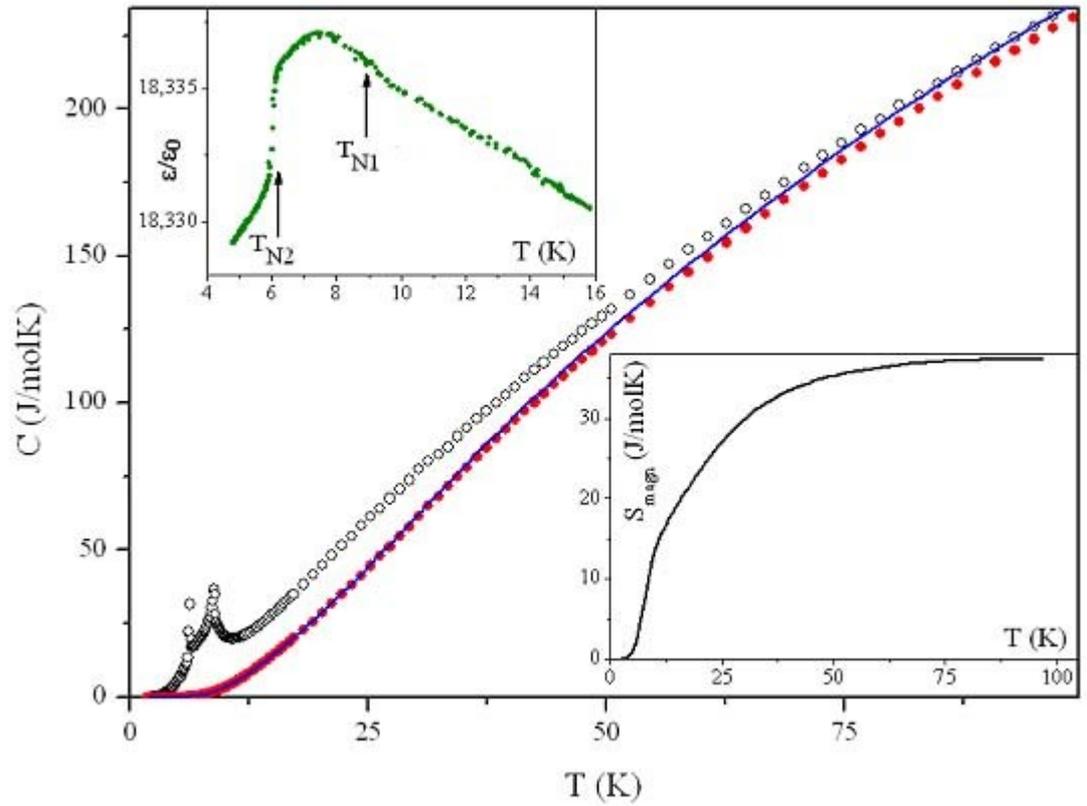

Fig. 5. The temperature dependences of specific heat in $Pb_3TeCo_3V_2O_{14}$ (open circles) and $Pb_3TeZn_3As_2O_{14}$ (solid circles). The solid line represents the result of scaling procedure employed for the estimation of magnetic contribution to the specific heat in $Pb_3TeCo_3V_2O_{14}$. Lower inset: the temperature dependence of the magnetic entropy in $Pb_3TeCo_3V_2O_{14}$. Upper inset: the temperature dependence of the dielectric constant in $Pb_3TeCo_3V_2O_{14}$.

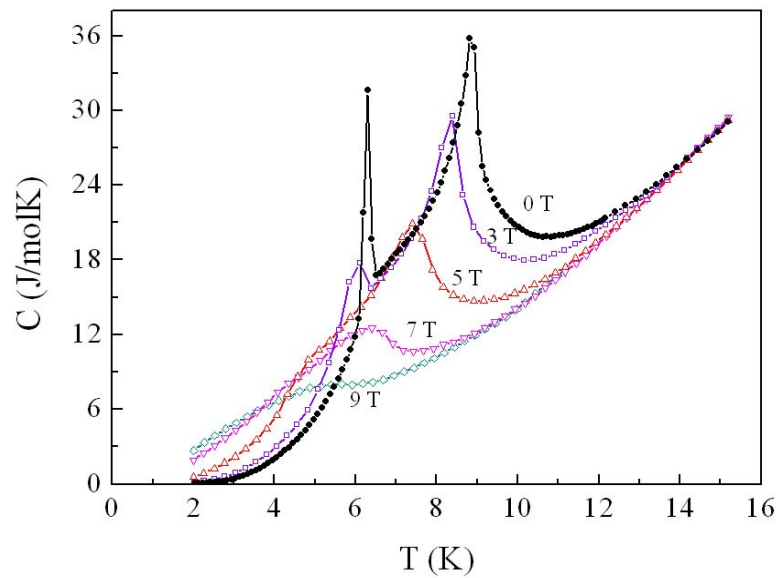

Fig. 6. The temperature dependences of specific heat $C(T)$ taken at various magnetic fields.

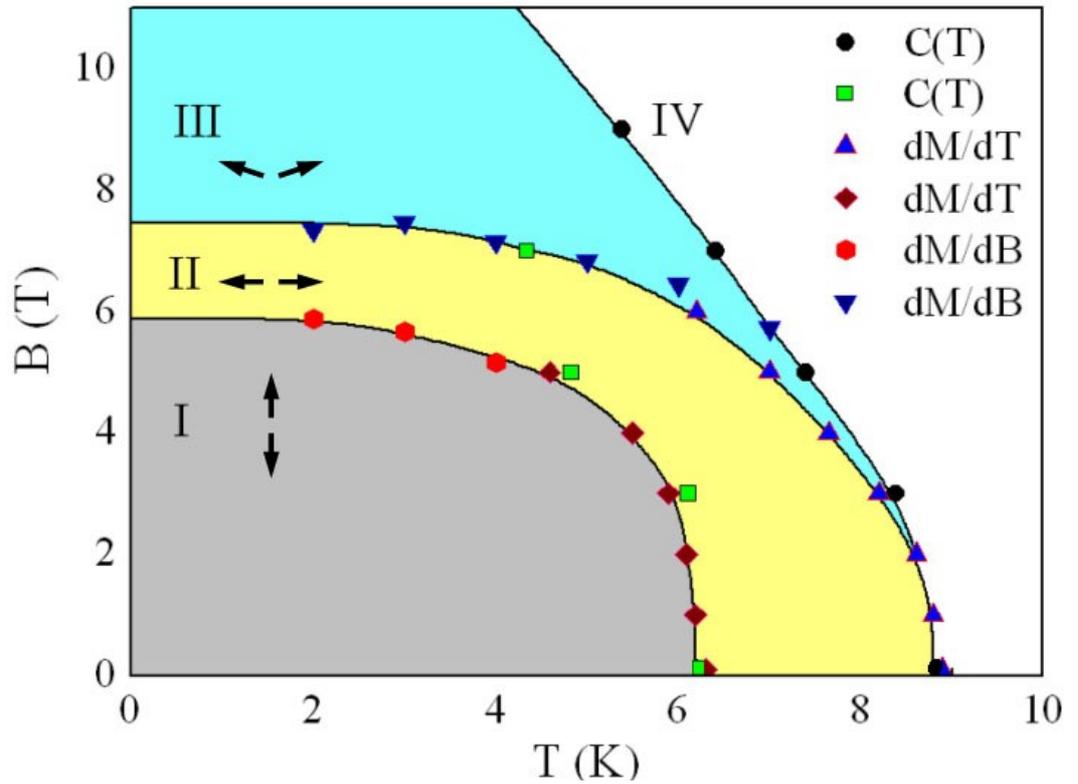

Fig. 7. The magnetic phase diagram of $Pb_3TeCo_3V_2O_{14}$ as defined from thermodynamic (magnetization $dM/dT$ and $dM/dB$ and specific heat $C(T)$) measurements. The phases I and II correspond to different antiferromagnetically ordered states. The phase III is a canted antiferromagnetic stateramagnetic state and the phase IV is a paramagnetic state.

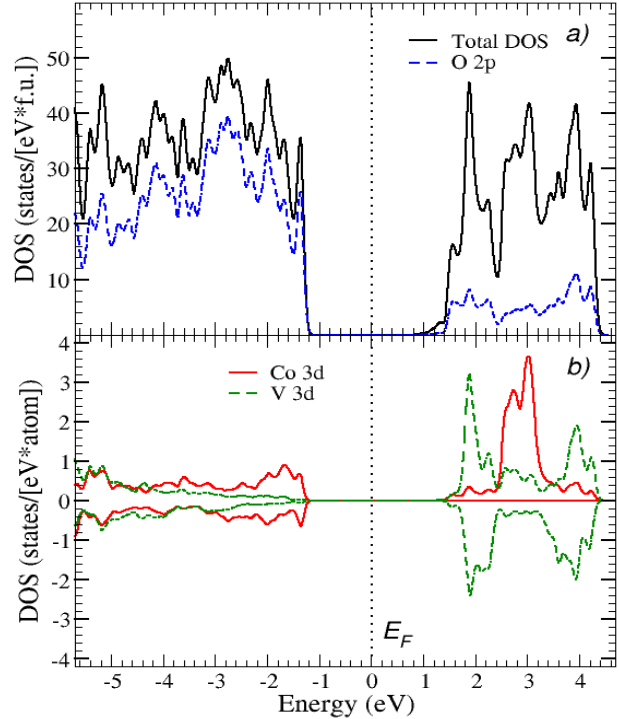

Fig. 8: The total and partial density of states of $Pb_3TeCo_3V_2O_{14}$ obtained in the GGA+U approximation for AFM-1 magnetic configuration (see Fig. 9). The Fermi energy is in zero.

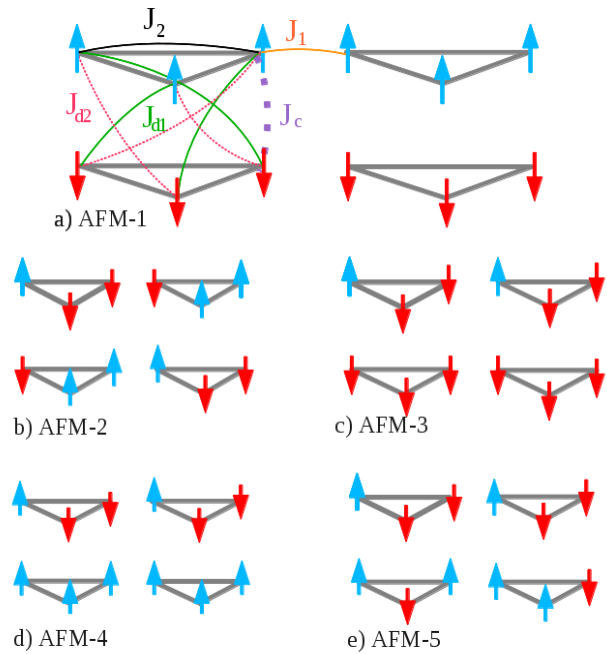

Fig. 9. The magnetic configurations used in the *ab initio* calculations to estimate different exchange integrals. Small Co triangles are shown in grey.

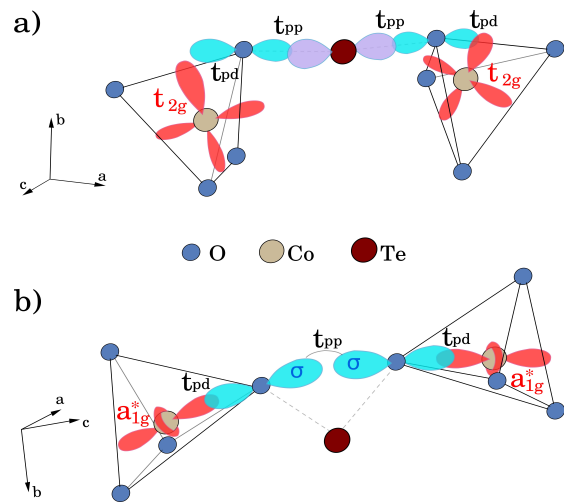

Fig. 10. The schematic representation of the superexchange paths for diagonal exchange couplings $J_{d1}$ and $J_{d2}$.